# A Novel Method on ISO 27001 Reviews: ISMS Compliance Readiness Level Measurement

Heru Susanto[12*], Mohammad Nabil Almunawar[1] and Yong Chee Tuan[1]

[1]FBEPS University of Brunei, Information System Group
[2]The Indonesian Institute of Sciences, Information Security & IT Governance Research Group

susanto.net@gmail.com, heru.susanto@lipi.go.id, nabil.almunawar@ubd.edu.bn

**Abstract:**

Security is a hot issue to be discussed, ranging from business activities, correspondence, banking and financial activities; it requires prudence and high precision. Since information security has a very important role in supporting activities of the organization, we need a standard or benchmark which regulates governance over information security. The main objective of this paper is to implement a novel practical approach framework to the development of information security management system (ISMS) assessment and monitoring software, called by I-SolFramework. System / software is expected to assist stakeholders in assessing the level of their ISO27001 compliance readiness, the software could help stakeholders understood security control or called by compliance parameters, being shorter, more structured, high precision and measured forecasting.

## 1. INTRODUCTION

Information security is not just a simple matter of having usernames and passwords (*Alan Calder and Setve Watkins, 2008*). Regulations and various privacy/data protection policy impose a raft of obligations to organization. Meanwhile, viruses, worms, hackers, phishers and social engineers threaten an organization on all sides. Research on information security area is extremely needed, especially in order to deal with the connectivity and cloud computing era. Access to high-quality, complete, accurate and up-to-date information is vital in supporting managerial decision-making process that leads to sound decisions. Thus, securing information system resources is extremely important to ensure that the resources are well protected. Regulations and various privacy/data protection policy impose a raft of obligations to organization. An organizational communication channel, which is using a network technology, such as intranet, extranet, internet, is a target for hackers in filtrated by [*figure 1*]. Although the development of IT security framework has gained much needed momentum in recent years, there continues to be a need for more writings on best theoretical and practical approaches to security framework development. Thus, securing information system resources is extremely important to ensure that the resources are well protected (*Chris Potter & Andrew Beard, 2008*).Furthermore, comprehensive and reliable information security controls reduce the organization's overall risk profile. ISO 27001 is the standard relating to Information Security Management System (ISMS). Companies or organizations obtained of ISO 27001Certificatemeaning a well-recognized for the security of information systems. Since information security has a very important role in supporting the activities of the organization, we need a standard as benchmark which regulates governance over information security. Several private and government organizations developed standards bodies whose function is to setup benchmarks, standards and in some cases, legal regulations on information security to ensure that an adequate level of security is preserved, to ensure resources used in the right way, and to ensure the best security practices adopted in an organization. There are several standards for IT Governance which leads to information security such as PRINCE2, OPM3, CMMI,P-CMM, PMMM, ISO27001, BS7799, PCIDSS, COSO, SOA, ITIL and COBIT. However, some of these standards are not well adopted by the organizations, with a variety of reasons. In this paper we will discuss the big five of ISMS standards, widely used standards for information security. The big five are ISO27001, BS 7799, PCIDSS, ITIL and COBIT. The comparative study conducted to determine their respective strengths, focus, main components and their level of adoption, concluded that ISO 27011 is most widely used standard in the world in information security area (*susanto, almunawar & yong, 2011b*). Unfortunately, many organizations find it difficult to implement ISO27001, including the obstacle when measuring the readiness level of an organizational implementation, which includes document preparation as well as various scenarios and strategy relating to information security (*susanto, almunawar & yong, 2011a*) and (*siponen & willison, 2009*).

## 2. SIX LAYER FRAMEWORK

This section we introduced new framework for approaching object and organization analyst, called by I-SolFramework, abbreviation from *I*ntegrated *Sol*ution for Information Security **Framework**. The framework consists of six layers component *[figure 1]*: organization, stakeholder, tools & technology, policy, culture, knowledge. Let us briefly introduced the basic elements of I-SolFramework, profile as illustrated (*susanto, almunawar & yong, 2011c*).

1. **Organization:** A social unit of people, systematically structured and managed to meet a need or to pursue collective goals on a continuing basis, the organizations associated with or related to, the industry or the service concerned (*BD, 2012*).

2. **Stakeholder:** A person, group, or organization that has direct or indirect stake in an organization because it can affect or be affected by the organization's actions, objectives, and policies(*BD, 2012*).

3. **Tools & Technology:** the technology upon which the industry or the service concerned is based. The purposeful application of information in the design, production, and utilization of goods and services, and in the organization of human activities, divided into two categories (1) Tangible: blueprints, models, operating manuals, prototypes. (2) Intangible: consultancy, problem-solving, and training methods (*BD, 2012*).

4. **Policy:** typically described as a principle or rule to guide decisions and achieve rational outcome(s), the policy of the country with regards to the future development of the industry or the service concerned (*BD, 2012*).

5. **Culture:** determines what is acceptable or unacceptable, important or unimportant, right or wrong, workable or unworkable. *Organization Culture:* The values and behaviors that contribute to the unique social and psychological environment of an organization, its culture is the sum total of an organization's past and current assumptions (*BD, 2012*).

6. **Knowledge:** in an organizational context, knowledge is the sum of what is known and resides in the intelligence and the competence of people. In recent years, knowledge has come to be recognized as a factor of production (*BD, 2012*).

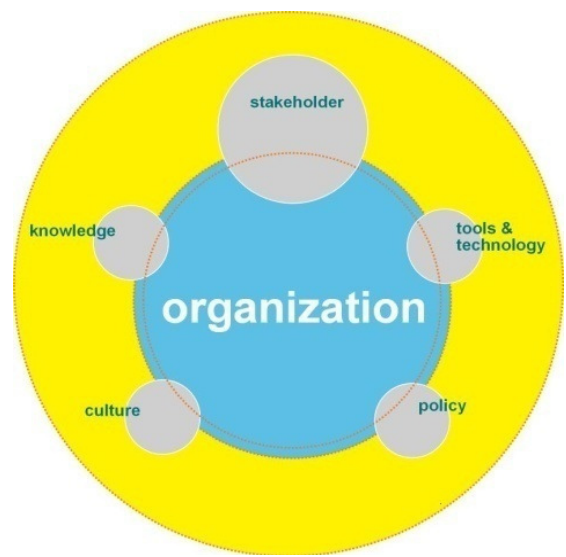

*Figure 1*.Integrated solution six domain framework

## 4. FRAMEWORK EQUATION

Determining the lowest level could be flexible, depending on the problems facing the object, might be up to $3^{rd}$, $4^{th}$, $5^{th}$...$N^{th}$ level. Formula works recursively; enumerate value from the lowest level, until the highest level of framework. Several variables indicated as level and contents position of framework indicator, where *k* as *control*, *j* as *section*, and *h* as a *top level*, details of these models are mention as follows.

$$(a) \rightarrow x_j = \sum_{k=1}^{n} \frac{[section]_k}{n}$$

$$x_j : control$$

$(a) \rightarrow x_j$ Indicate value of control of ISO which is resulting from *sigma* of section(s) assessment, divided by number of section (s) contained on the lowest level.

$$(b) \rightarrow x_i = \sum_{j=1}^{n} \frac{[control]_j}{n}$$

$$x_i : domain$$

$(b) \rightarrow x_i$ Stated value of domain of ISO which is resulting from *sigma* of control(s) assessment, divided by number of control(s) contained at concerned level.

After *(a), (b),* defined, then the next step is substituted of mathematical equations mentioned, into new comprehensive modelling notation in a single mathematical equation, as follows.

$$x_h = \sum_{i=1}^{n} \frac{[control]_i}{n}$$

$$x_h = \sum_{i=1}^{n} \frac{[b]_i}{n}$$

$$x_h = \sum_{i=1}^{n} \frac{\left[\sum_{j=1}^{n} \frac{[section]_j}{n}\right]_i}{n}$$

*So that for six layer, or we called it by top level, equation will be:*

$$x_h = \sum_{i=1}^{6} \frac{\left[\sum_{j=1}^{n} \frac{\left[\sum_{k=1}^{n} \frac{[section]_k}{n}\right]_j}{n}\right]_i}{6}$$

*Where;   k=section; J=control; I=domain (organization, stakeholder, tools & technology, policy, knowledge, and culture).*

The algorithm is considered to be reliable and easy implementing in analysing such problem (*susanto, almunawar&tuan, 2011d*), emphasized on divided problems into six layers as the initial reference in measuring object. Indicated that layer with a weak indicator has a high priority for improvement and refinement within organization as a whole. In the manner of the six layer framework, I-SolFramework, analysis could be works easily and simply observe.

## 5. SOFTWARE DESIGN AND FEATURES

The ISO recommendations are associated with two levels of security protection: a basic level that considers essential security controls; and an extended level that extends the essential controls in order to provide additional security protection (*Kosutic, 2010*). It starts with the "21 essential security controls" of ISO 27001, which give the basic standard requirements of information security management. Controls are mapped on these domains and subsequently refined into "246 simple and easily comprehended elements". These elements are subject to be reviewed and validated by specialized persons working on the field.

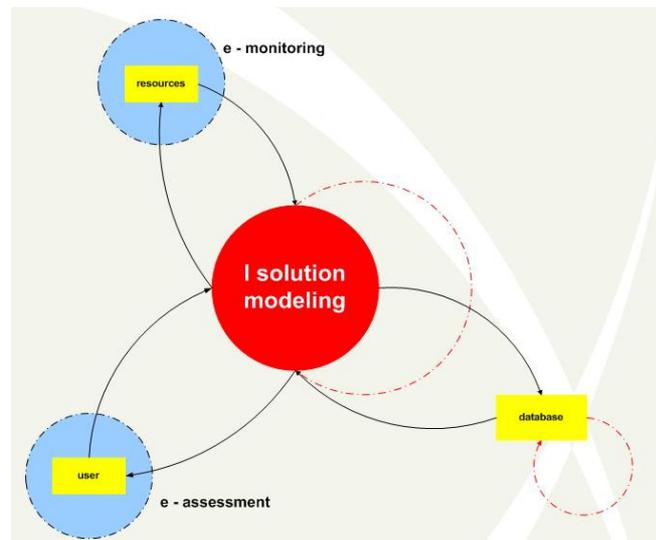

*Figure 2*. I-Solution Modeling Data flow diagram level 0

In general, i-Solution Modelling software consists of two major subsystems of e-assessment and e-monitoring [*figure 2*]. E-assessment to measure ISO 27001 parameters based on the proposed framework [*figure 1*] with 21 controls [*figure 3*]. Software is equipped with a login system, as the track record of the user, as function as to knows how many times try the experiment in assessing the organization so it can be determined patterns of assessment.

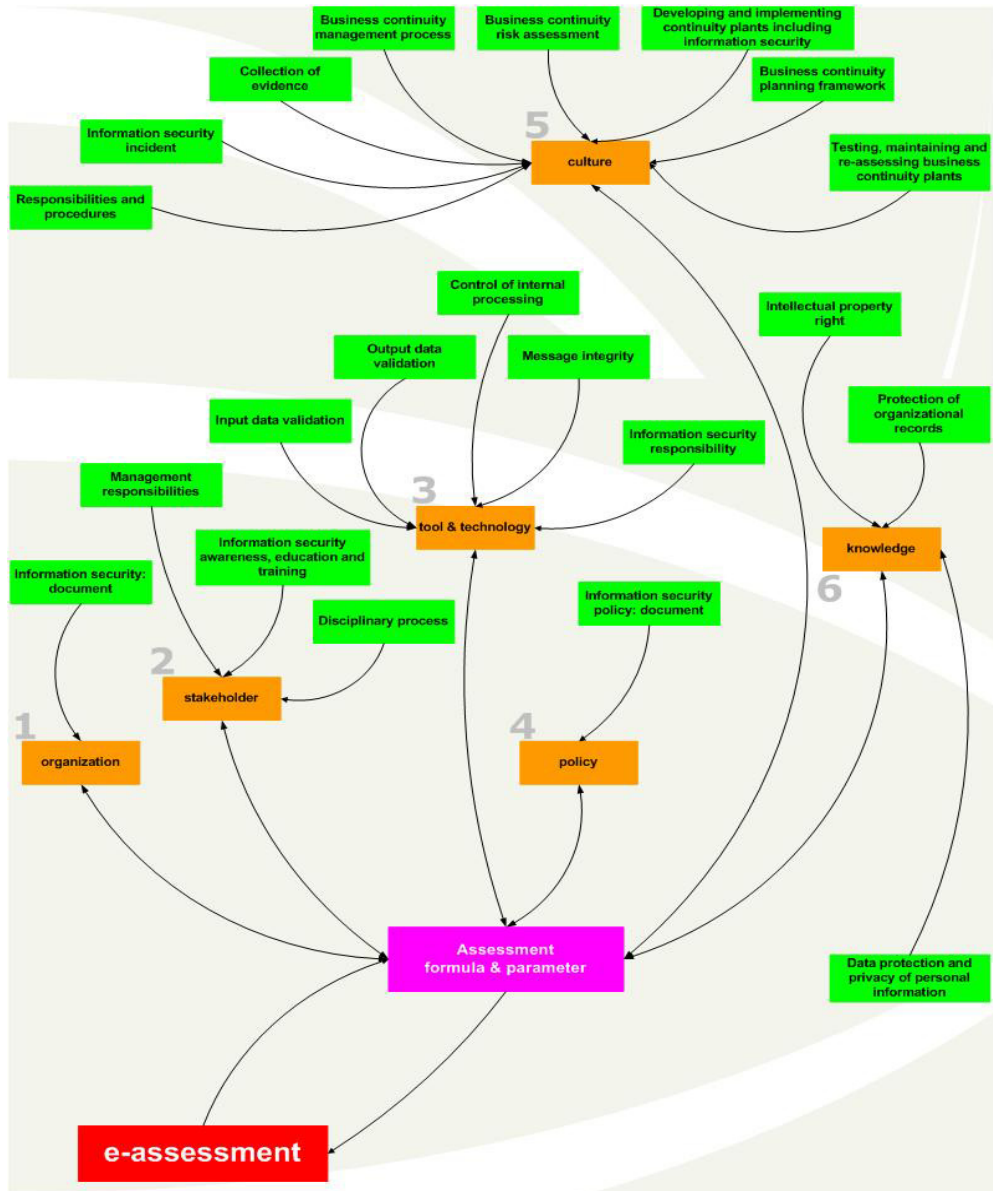

*Figure 3*. I-Solution Modeling Data flow diagram level 1

E-assessment utility is validating of the ISO 27001 parameters, through user interface provided by the system, follows i-solution framework rules that divided and segmented ISO 27001 essential controls into six main domains [*figure 2*].

In the main form, software is featuring by three main tabs as function as:

- ***Tab Assessment***
  On this sub form, user is prompted to entering an achievements value based onISO27001 parameters, called by assessment issues. Level of assessment set out in range of 5scales;
  - *0 = not implementing*
  - *1= below average*
  - *2=average*
  - *3=above average*
  - *4=excellent*

As a measurement example, we described it in details to several steps of parameters assessment as follows [*figure 6*]:
  1. *Domain*: "**Organization**"
  2. *Controls*: "**Organization of information security: Allocation of Information Security Responsibilities**"
  3. *Assessment Issue*: "**Are assets and security process Cleary Identified?**"
  4. Then stakeholders should analogize ongoing situation, implementation and scenario in organization, and benchmark it to the security standard level of assessment as reference standard.

*Figure 4. Assessment form*

- ***Tab Histogram***
  Histogram showed us details of the organization's achievement and priority. Both statuses are important in reviews of strongest and weakness point on an organization current achievement.

As indicated is the system, *"Achievement"* declared the performance of an organization as final result of the measurement by validated by proposed framework. Then another term is *"Priority"* indicated the gap between ideal values with achievement value. "Priority" and" achievement" showed inverse relationship. If achievement is high, then domain has a low priority for further work, and conversely, if achievement is low, then the priority be high [*figure 5*].

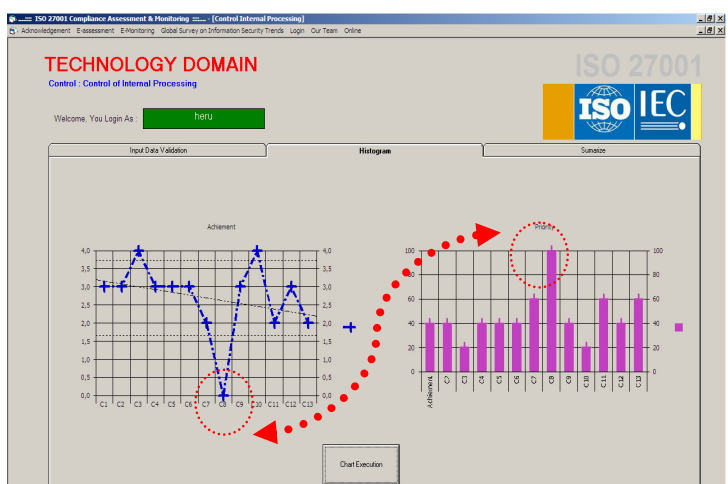

*Figure 5. Final result view on histogram style*

- *<u>Tab Summarize</u>*

    Submenu summarize has a feature on provided a user to analyse the results of their estimates. Some of the assessment criteria are displayed, in addition system advice is prepared, and it gives advice based on the previous assessment. Four main features are:

    - Final result out of 4 scale
    - Final result out of 100 %
    - Final predicate of assessment result (not implementing, below average, average, above average, excellent)
    - Advice from the software regarding their final achievement, in which point their strongest area and also their weakness area.

By marking estimated performance values for each parameter, as assessment and forecasting approach as well, stakeholders have a comprehensive overview achievement on their readiness level. In some cases, to assess their organization's readiness level, stakeholders needed more than once of experiment, marked by a significant increase in the final result grade of each experiment. One first experiments on stakeholder assessment called by first trying and training course, time required for each experiment was 30 minutes to 60 minutes, this achievement represents a significant contribution to an organization in understanding the ISO27001 controls, clause and assessment issues as well, than normally step which is need between 12months – 24months (*iso27001security.com*).

## 7. AN ILLUSTRATIVE MEASUREMENT

An illustrative example is presented to delineate usability level of its approach. Each question of the refined simple elements, a value associated with the example is given. *Table 2* summarizes the results of all domains together with their associated controls (*susanto, almunawar & yong, 2011c*) & (*alfantookh, 2009*), based on I-SolFramework measurements approaches. The results given are illustrated in the following figures. *Figure 9* illustrates the state of five essential controls of the "tools &technology" domain. *Figure 8* represents the condition of 21 essential controls of standards by table and also *Figure 10* stated overall condition of 21 essential controls in histogram style.The overall score of all domains is shown in the Table to be "2.66 points". The domain of the "policy" scored highest at "4", and the domain of the "knowledge" scored lowest at "2". Ideal and priority figures are given to illustrate the strongest and weaknesses in the application of each control (*figure 6&8*).

*Figure 6. Final result view on summarize style*

Computer Science Journal                                   Volume 2, Issue 1, April 2012

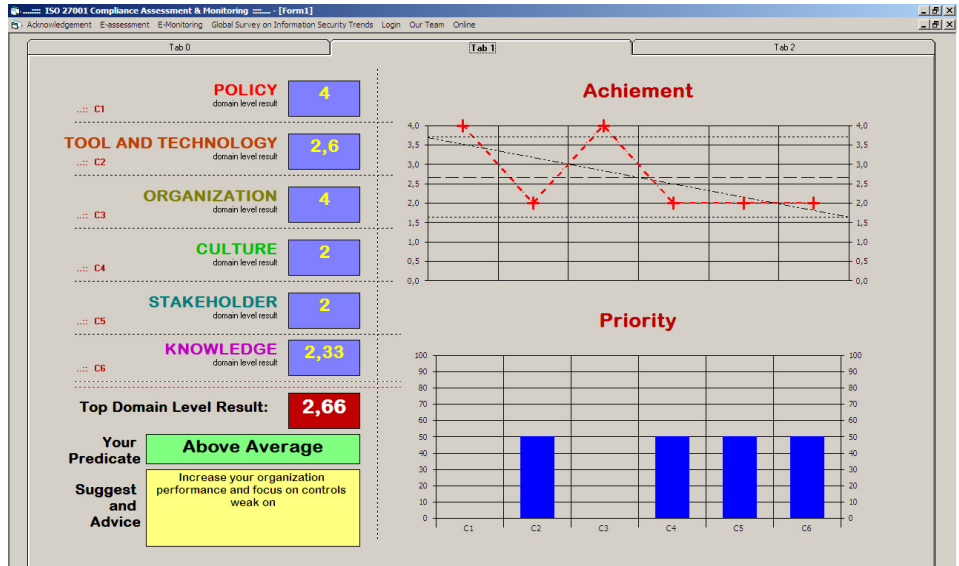

*Figure 7*. *Six domain final result view on histogram style*

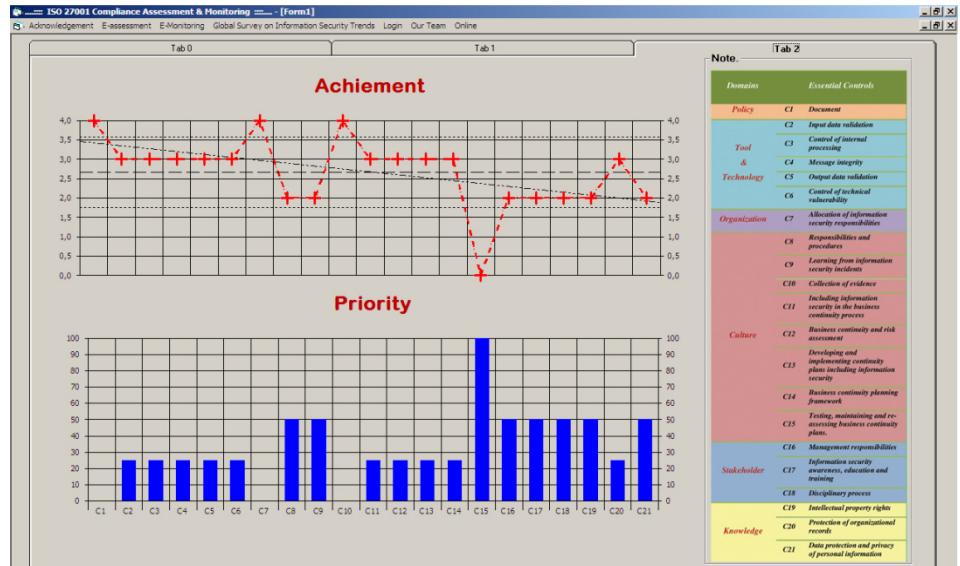

*Figure 8*. *21-essential controls final result view on histogram style*

## 8. CONCLUSION REMARKS

I-Solution modelling is software which has new paradigm framework, to make assessments and monitoring, since securing and maintaining information from parties who do not have authorization to access such information is crucial priority. It is expected to provide solutions to solved obstacles, challenges and difficulties in understanding standard term and concept, as well as assessing readiness level of an organization towards implementation of ISO 27001forinformation security. On trials conducted, user can perform a test of his organization assessment within 30-60minutes; in expected perhaps it could reduce time in understanding and assessing set of standard parameters leads obtain certification of information security.

## AUTHORS

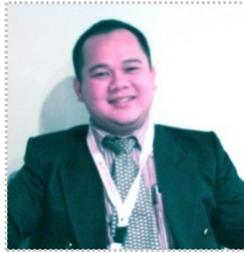
**HeruSusanto** is a researcher at The Indonesian Institute of Sciences, Information Security & IT Governance Research Group, also was working at Prince Muqrin Chair for Information Security Technologies, King Saud University. He received BSc in Computer Science from Bogor Agriculture University, in 1999 and MSc in Computer Science from King Saud University, and nowadays as a PhD Candidate in Information Security System from the University of Brunei.

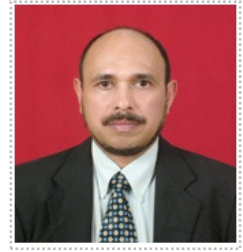
**Mohammad Nabil Almunawar** is a senior lecturer at Faculty of Business, Economics and Policy Studies, University of Brunei Darussalam. He received master Degree (MSc Computer Science) from the Department of Computer Science, University of Western Ontario, Canada in 1991 and PhD from the University of New South Wales (School of Computer Science and Engineering, UNSW) in 1997. Dr Nabil has published many papers in refereed journals as well as international conferences. He has many years teaching experiences in the area computer and information systems.

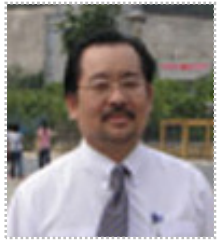
**Yong Chee Tuan** is a senior lecturer at Faculty of Business, Economics and Policy Studies, University of Brunei Darussalam, has more than 20 years of experience in IT, HRD, e-gov, environmental management and project management. He received PhD in Computer Science from University of Leeds, UK, in 1994. He was involved in the drafting of the two APEC SME Business Forums Recommendations held in Brunei and Shanghai. He sat in the E-gov Strategic, Policy and Coordinating Group from 2003-2007. He is the vice-chair of the Asia Oceanic Software Park Alliance.